\documentclass{article}
\title{Exact Solution of a Linear Wave Equation \\ in Cosmological General Relativity}
\author{Firmin J. Oliveira $^\dagger$}

\begin{document}

\maketitle

\begin{center}
{$^\dagger$ Joint Astronomy Centre, \label{fn:affil}
           Hilo, Hawai{`}i, U.S.A. \\
  Email: {firmin@jach.hawaii.edu} \\
  Address:  {660 N. A{`}ohoku Place, Hilo, Hawai{`}i, U.S.A. 96720, \\
            Telephone: 808-969-6539, \\
            Fax: 808-961-6516.}}
\end{center}

\begin{abstract}

A linear second order wave equation is presented based on cosmological general relativity,
which is a space-velocity theory of the expanding Universe. The wave equation is shown to
be exactly solvable, based on the Gaussian hypergeometric function.
\end{abstract}

\section{Introduction}

In this paper a linear second-order wave equation is presented based on the space-velocity
theory of cosmological general relativity (CGR) of Carmeli\cite{carmeli-0}. A spherically
symmetric metric is used with comoving coordinates. The equation is separated by variables
and exact solutions are found for all ensuing equations. The interesting radial equation
is transformed into a Gaussian hypergeometric equation for which solutions are well known.

Note that this paper is more rigorous than \cite{oliveira-1}, although the section on the
derivation of the wave equation is essentially the same.

\section{The Wave Equation \label{sect:wave}}

The CGR spherically symmetric, comoving space-velocity metric (\cite{carmeli-2}, Eq. (A.5))
is defined by the line element
\begin{equation}
  ds^{2} = \tau^{2} dv^{2} - e^{\mu} dr^{2}
           - Q^{2} \left( d{\theta}^{2} + sin^{2}(\theta) d{\phi}^{2} \right) \, , \label{eq:sph-sym-metric}
\end{equation}
where coordinate $(v)$ is the radial velocity of expansion of the universe, and $\tau$ is the Hubble-Carmeli time
constant, its value is $\tau=12.486 \,{\rm Gyr}\,$ (\cite{carmeli-2}, Eq. (A.66), p. 138).
The functions $\mu$ and $Q$ are dependent only on the velocity $(v)$ and the radial coordinate $(r)$.
$(\theta,\phi)$ are the usual spherical coordinates. From Eq. \ref{eq:sph-sym-metric} the non-zero
elements of the metric $g_{\mu \nu}$ are
\begin{eqnarray}
& &  g_{0 0} = 1\, , \\ 
& &  g_{1 1} = -e^{\mu} \, , \\
& &  g_{2 2} = -Q^{2}\, , \\
& &  g_{3 3} = -Q^{2} sin^{2}(\theta) \, .
\end{eqnarray}

A postulate of this cosmological theory is that the metric $g_{\mu \nu}$ satisfies the Einstein field equations
\begin{equation}
  G_{\mu \nu}  =  R_{\mu \nu} - \frac{1}{2} g_{\mu \nu} R =  \kappa T_{\mu \nu}  \, , \label{eq:Einstein-eqn}
\end{equation}
where $ R_{\mu \nu}$ is the Ricci tensor, $R=g^{\alpha \beta} R_{\alpha \beta}$ is the scalar curvature, 
$T_{\mu \nu}$ is the momentum-energy tensor,
$\kappa = 8 \pi {\rm G} / (c^{2} \tau^{2})\,$, ${\rm G}$ is Newton's constant and $c$ is the speed of
light in vacuo. (Compare this to the Einstein equation in General Relativity theory where
$\kappa = 8 \pi {\rm G} / c^{4}$.) The momentum-energy tensor of a perfect fluid
 (\cite{carmeli-2}, Eq. (A.10)) is 
\begin{equation}
  T_{\mu \nu} = \rho_{eff}\, u_{\mu} u_{\nu}
               + p  \left( u_{\mu} u_{\nu} -  g_{\mu \nu} \right) \, ,
\end{equation}
where the effective mass density $\rho_{eff} = \rho - \rho_{c}$, where $\rho$ is the average mass density
of the Universe and $\rho_{c}$ is the  critical mass density, a {\em constant} in CGR given by 
 $\rho_{c} = 3/(8 \pi {\rm G} \tau^{2})$. Also, $p$ is the pressure, and $u^{\mu}$ is the four-velocity
\begin{equation}
u^{\alpha} = u_{\alpha} = \left(1, 0, 0, 0 \right). 
\end{equation}

The solution for Eq. \ref{eq:Einstein-eqn} was derived (\cite{carmeli-2}, Appendix A.4) with the results,
   \begin{eqnarray}
      Q & = & r\, , \label{eq:nu_solution} \\
  \nonumber \\
      e^{\mu} & = &  e^{\mu(r)} = \frac{1}{1 + f(r)} \, , \label{eq:mu_solution} \\
  \nonumber \\
      f(r) & = & \frac{1 - \Omega}{c^{2} \tau^{2}} r^{2} \, , \label{eq:fr} \\
  \nonumber \\
      p & = & \frac{c}{\tau} \frac{\left( 1 - \Omega \right)}{8 \pi {\rm G}} \, ,
   \end{eqnarray}
where the density parameter $\Omega = \rho / \rho_{c}$.

A linear second order wave equation is obtained from the dual \cite{ohara} of Eq. \ref{eq:sph-sym-metric}
in the form of the d`Alembertian for space-velocity
\begin{eqnarray}
 & &  \frac{\partial^{2}{\Psi}}{\partial s^{2}} =
  \left\{ \frac{\partial^{2}}{\partial{\left(\tau v\right)^{2}}}
        -e^{-\mu(r)} \frac{1}{r^{2}}\frac{\partial}{\partial{r}}
             \left( r^{2}\frac{\partial}{\partial{r}} \right) \right. \nonumber \\
\nonumber \\
  & & - \left. \frac{1}{r^{2}}
         \left[ \frac{1}{sin(\theta)} \frac{\partial}{\partial{\theta}}
                \left( sin(\theta) \frac{\partial}{\partial{\theta}} \right)
      + \frac{1}{sin^{2}(\theta)} \frac{\partial^{2}}{\partial{\phi^{2}}} \right] \right\} \Psi \, .  \label{eq:Psi_wave}
\end{eqnarray}

In CGR the condition for the expansion of the universe is defined by setting $ds=0$.
For the wave equation, it is assumed that the expansion of the universe corresponds to setting 
 $\partial^{2}{\Psi} / \partial{s^{2}} = 0$. With this condition,
Eq. \ref{eq:Psi_wave} becomes
\begin{eqnarray}
& & \frac{1}{\tau^{2}} \frac{\partial^{2}{\Psi}}{\partial{v^{2}}} =
     e^{-\mu(r)} \frac{1}{r^{2}} \frac{\partial}{\partial{r}} \left( r^{2} \frac{\partial{\Psi}}{\partial{r}} \right)
                \nonumber \\
  \nonumber \\
& &        +\frac{1}{r^{2}}
  \left[ \frac{1}{sin(\theta)} \frac{\partial}{\partial{\theta}}
                \left( sin(\theta) \frac{\partial{\Psi}}{\partial{\theta}} \right)
                           + \frac{1}{sin^{2}(\theta)} \frac{\partial^{2}{\Psi}}{\partial{\phi^{2}}} \right]
                   \, .   \label{eq:stat-wave-sq3}
\end{eqnarray}

\section{Solution of the Wave Equation}

Solve Eq. \ref{eq:stat-wave-sq3} by a separation of variables. Assume
\begin{equation}
 \Psi(v,r,\theta,\phi) = \Psi_{0}(v)\, \Psi_{1}(r)\, \Psi_{2}(\theta)\, \Psi_{3}(\phi) \, .
\end{equation}
By the well known process, the solutions of the equations for the $(v)$ and $(\theta, \phi)$
components are readily obtained with the result
\begin{eqnarray}
& &  \Psi_{0}(v) = e^{\pm i D \tau v} \, ,  \label{eq:Psi_0} \\
 \nonumber \\
& &  \Psi_{2}(\theta) \Psi_{3}(\phi) = P^{m}_{l}(cos(\theta)) e^{+ i m \phi} \, , \label{eq:Theta_Phi_solution}
\end{eqnarray}
where $D$ is a constant of integration called the {\em intrinsic curvature},
and where $P^{m}_{l}(cos(\theta)) $ are the associated Legendre functions with
\begin{eqnarray}
  & &  l = 0, 1, 2, 3, \ldots \, ,  \\
\nonumber \\
  & &  m = -l, -(l-1), \ldots, 0, \ldots, l-1, l \, .
\end{eqnarray}
This leaves the radial equation. Substituting for $e^{\mu(r)}$ from
Eqs. \ref{eq:mu_solution} and \ref{eq:fr}, the equation for $\Psi_{1}(r)$ is,
\begin{eqnarray}
& &     r^{2} \left( 1 + A r^{2} \right) \frac{d^{2}\Psi_{1}(r)}{dr^{2}}
     + 2 r \left( 1 + A r^{2} \right) \frac{d\Psi_{1}(r)}{dr} \nonumber \\
  \nonumber \\
& &     + \left[ D^{2} r^{2} - l \left( l + 1 \right) \right] \Psi_{1}(r) = 0 \, ,  \label{eq:R-wave-eq} 
\end{eqnarray}
where
\begin{eqnarray}
  & &  A  = \frac{1 - \Omega}{c^{2} \tau^{2}} \, , \label{eq:A} \\
\nonumber \\
  & & 0 \le \Omega < \infty \, .
\end{eqnarray}

This equation is transformed (\cite{polyanin}, Sect. 2.1.2-6, Eq. 194)
 in terms of $U(x)$ by defining
\begin{equation}
  x = -A\, r^{2} \, , \label{eq:x}
\end{equation}
and the ansatz
\begin{equation}
  \Psi_{1}(x) = x^{q} U(x) \, , \label{eq:Psi=si_U}
\end{equation}
where $q$ is a solution to the quadratic equation
\begin{eqnarray}
 q^{2} - \frac{5}{2} q - \frac{1}{4} l \left( l + 1 \right) = 0 \, .
\end{eqnarray}
Apply this tranformation to Eq. \ref{eq:R-wave-eq} to obtain the Gaussian hypergeometric equation
(\cite{polyanin}, Sect. 2.1.2-5, Eq. 171)
\begin{eqnarray}
  x \left( x - 1 \right) \frac{d^{2}U(x)}{dx^{2}}
     + \left[\left( \alpha + \beta + 1 \right) x - \gamma \right] \frac{d U(x)}{dx}
     + \alpha\, \beta\, U(x) = 0 \, , \label{eq:gauss-hyper}
\end{eqnarray}
where
\begin{eqnarray}
& &  q = \frac{5}{4} \pm \frac{1}{2} \sqrt{\frac{25}{4} + l \left( l + 1 \right)} \, , \label{eq:k_eqn} \\
  \nonumber \\
& &  \alpha = q - \frac{5}{4} \pm \frac{1}{2} \sqrt{\frac{25}{4} - \frac{D^{2}}{A}} \, , \label{eq:beta_eqn} \\
  \nonumber \\
& &  \beta = 2 q - \frac{5}{2} - \alpha \, , \\
  \nonumber \\
& &  \gamma = 2 q - \frac{3}{2} \, . \label{eq:gamma_eqn}
\end{eqnarray}

To prove that $\gamma$ is not an integer, substitute for $q$ from Eq. \ref{eq:k_eqn}
into Eq. \ref{eq:gamma_eqn} and obtain
\begin{eqnarray}
  25 = 4 \left( \gamma - 1 \right)^{2} - 4\, l \left( l + 1 \right)  \, ,  \label{eq:contradiction}
\end{eqnarray}
which is false for integers, since the l.h.s. is odd while the r.h.s. is even.

For $\gamma \ne 0, -1, -2, -3, \ldots$, a solution to Eq. \ref{eq:gauss-hyper} is the hypergeometric series
\begin{eqnarray}
& &  U(x) = F(\alpha, \beta, \gamma;x) =
    1 + \sum^{\infty}_{k=1}{\frac{\left( \alpha\right)_{k} \left( \beta\right)_{k} }
                               {\left( \gamma \right)_{k}}
                          \frac{x^{k}}{k !}} \, , \label{eq:U(x)=F(a,b,c;x)} \\
 \nonumber \\
& &  \left(\alpha \right)_{k} =
    \alpha \left( \alpha + 1 \right)  \left( \alpha + 2 \right) \cdots \left( \alpha + k - 1 \right) \, ,
\end{eqnarray}
which, {\em a fortiori}, is convergent for
\begin{equation}
 \mid x \mid < 1 \, . \label{eq:x_converge}
\end{equation}

For $\gamma$ not an integer, and for $C_{1}$ and $C_{2}$ arbitrary constants, the general solution
of Eq. \ref{eq:gauss-hyper} is
\begin{equation}
  U(x) = C_{1} F(\alpha, \beta, \gamma; x)
    + C_{2}\, x^{1-\gamma} F(\alpha - \gamma + 1, \beta - \gamma + 1, 2 - \gamma; x)  \, . \label{eq:U(x)_general}
\end{equation}

Consider the radial wave function $\Psi_{0}(x) = x^{q} U(x)$, for $\mid x \mid \, < 1$.
As $x \rightarrow 0$, only the first term in each series is relevant in Eq. \ref{eq:U(x)_general},
\begin{eqnarray}
  \Psi_{0}(x \rightarrow 0) & \approx & x^{q} \left (C_{1} + C_{2} x^{1 - \gamma} \right) \, , \\
\nonumber \\
   & \approx & C_{1} x^{q} + C_{2} x^{5/2 - q} \, .  \label{eq:Psi0(0)}
\end{eqnarray}
where substitution for $\gamma$ was made from Eq. \ref{eq:gamma_eqn}.  If $q$ is positive then the
second term of Eq. \ref{eq:Psi0(0)} will diverge as $x \rightarrow 0$ for $5/2 < q$,
which requires that $C_{2} = 0$ for $5/2 < q$. If $q$ is negative then the first term of Eq. \ref{eq:Psi0(0)}
will diverge as $x \rightarrow 0$ unless $C_{1} = 0$.  If $q=0$ there is no divergence.

\section{Some Physical Aspects}

From Eqs. \ref{eq:A}, \ref{eq:x} and \ref{eq:x_converge},
\begin{equation}
\mid x = -\frac{\left( 1 - \Omega \right)  r^{2}}{c^{2} \tau^{2}} \mid < 1 \, ,
\end{equation}
which implies
\begin{eqnarray}
& &  0 \le r < R_{\Omega} \, , \\
\nonumber \\
& &  R_{\Omega} = \sqrt{\mid R^{2}_{\Omega} \mid} \, , \\
\nonumber \\
& &  R^{2}_{\Omega} = -\frac{c^{2} \tau^{2}}{1 - \Omega} \, .
\end{eqnarray}
The radius of convergence $R_{\Omega}$ is identical to the radius of curvature of the cosmological models
for the expanding universe described in (\cite{behar-1}, Eqs. 2.4, 2.5, 5.8a and 5.8b).
The conclusion is that the wave function solution $\Psi(v,r,\theta,\phi)$ found here is valid for the entire
expanding universe of radius $R_{\Omega}$.

An expression for the intrinsic curvature is obtained by eliminating $q$ between
Eqs. \ref{eq:k_eqn} and \ref{eq:beta_eqn},
\begin{equation}
      D = \pm \frac{\sqrt{\Omega - 1}}{c\, \tau}
     \left[ \left( 2\, \alpha  \pm \sqrt{\frac{25}{4} + l(l+1)} \right)^{2}  - \frac{25}{4} \right]^{1/2}
       \label{eq:D_eqn}
\end{equation}
Notice that $D$ can be real or imaginary, depending on $\Omega$, $\alpha$ and $l$,
which means $\Psi_{0}(v)$ can be a sinusoidal wave or a decaying exponential.

\vskip 1cm

  The author is grateful to the Joint Astronomy Centre, Hilo, Hawai{`}i for many years of employment and support,
  and to Professor Moshe Carmeli for his theories and communications.

\end{document}